\documentclass[runningheads]{llncs}

\usepackage{graphicx}
\usepackage{amsmath}
\usepackage{amsfonts}
\usepackage{comment}

\begin{document}
\title{Feedback Assisted Adversarial Learning to Improve the Quality of Cone-beam CT Images}

\titlerunning{Feedback Assisted Adversarial Learning}
\author{Takumi Hase\inst{1} \and
Megumi Nakao\inst{1} \and
Mitsuhiro Nakamura\inst{2} \and
Tetsuya Matsuda\inst{1}}

%\authorrunning{M. Nakao et al.}
\institute{Graduate School of Informatics, Kyoto University, Kyoto, Japan.
\email{megumi@i.kyoto-u.ac.jp}\\
\and
Graduate School of Medicine, Human Health Sciences, Kyoto University, Kyoto, Japan.\\}

\maketitle

\begin{abstract}
Unsupervised image translation using adversarial learning has been attracting attention to improve the image quality of medical images. However, adversarial training based on the global evaluation values of discriminators does not provide sufficient translation performance for locally different image features. We propose adversarial learning with a feedback mechanism from a discriminator to improve the quality of CBCT images. This framework employs U-net as the discriminator and outputs a probability map representing the local discrimination results. The probability map is fed back to the generator and used for training to improve the image translation. Our experiments using 76 corresponding CT-CBCT images confirmed that the proposed framework could capture more diverse image features than conventional adversarial learning frameworks and produced synthetic images with pixel values close to the reference image and a correlation coefficient of 0.93.
\end{abstract}

\section{Introduction}
Cone-beam computed tomography (CBCT) has been increasingly used for 3D imaging in clinical medicine. Because the device is compact and movable, 3D images can be acquired from the patient during surgery or radiotherapy. Conversely, CBCT, in which X-rays are irradiated in a conical shape and signals are obtained by a two-dimensional detector, is more susceptible to scattering than CT. Signals cannot be obtained from a certain range of directions because of the physical constraints of the rotation angle. These differences in signal characteristics result in a wide range and variety of artifacts in the reconstructed image \cite{Tang18}. Due to the effect of these artifacts, the CBCT images' pixel values are inaccurate compared to CT images, which is a factor that reduces the image quality and accuracy of medical image analysis. %If the quality of CBCT images can be improved by utilizing the intrinsic relationships between biological tissues and pixels in CT images, and statistical knowledge, the expansion of their use in surgical support and radiotherapy can be expected, which could lead to the creation of new diagnosis and therapy processes.

There are many examples of research into using deep learning to improve the quality of medical images \cite{Yi19}. Attempts have been made to improve the quality of CBCT images by scattering correction \cite{Zollner17}. Several methods are based on supervised learning and require CT images that perfectly match the anatomical structures in the CBCT images. However, it is difficult to obtain a set of images with perfectly matched structures for the same patient, and the opportunities for using this technique are limited. To address this issue, unsupervised learning that does not assume a one-to-one correspondence between target images has been widely studied, especially by applying Generative Adversarial Networks (GANs) \cite{Zhu17}. The CycleGAN is reportedly effective in reducing dental metal artifacts in CT images \cite{Nakao20}\cite{Nakamura21}, and artifacts in craniocervical CBCT images \cite{Liang19}.

%一方，CBCT画像におけるアーチファクトは，複数の発生要因から生じるために多様な画像特徴を示し，また画像内の広範囲に渡って存在する．このため，従来研究で提案された学習の枠組みや損失関数の正則化\cite{Kida20}\cite{Hase21} によっても十分な画質改善が達成されない場合が存在する．この画像変換性能の限界の一要因として，従来のGANやCycleGANにおける識別器が目標画像か否かを大域的な評価値によってのみ評価している点が挙げられる．大域的な評価では局所的に異なる特徴を有する画像特徴に対して十分な変換性能が得られないことが報告されている\cite{Edgar20}．局所的に異なる画像特徴を捉えつつ，多様な画像生成を実現する敵対的学習，画像変換の達成は依然課題である．

Artifacts in CBCT images exhibit various image features and are present over a wide area of the image. Therefore, cases occur in which existing adversarial training and regularization of the loss function \cite{Kida20}\cite{Hase21} inadequately improve image quality. A reason for this limitation on image translation performance is that the discriminator in conventional GANs evaluates whether or not an image is a target image based on a global evaluation value alone. Reportedly, a global evaluation does not provide sufficient translation performance for image features with different local characteristics \cite{Edgar20}. To obtain adversarial learning and image translation that generates a variety of images while capturing locally different image features remains a challenge.

%本研究では，CBCT画像の画質改善を目的とした新たな教師なし学習の枠組みとして，識別器からのフィードバック機構を備えた敵対的学習を提案する．従来のGANでは，識別器が出力する大域的な識別結果に基づいて敵対的学習が行われてきたが，提案方法では確率マップを出力する識別器へと拡張して用いる．ここで，確率マップとは，画素単位で入力画像が目標画像である確率を算出して格納した画像であり，生成画像と目標画像間に見られる画像特徴の差に関する局所的な評価を得ることを目的としている．また従来のGANでは，識別器による識別結果は損失関数における評価にのみ用いられてきた．提案する敵対的学習の枠組みでは，識別器が出力する確率マップを生成器へ追加的に入力するフィードバック機構を導入する．生成器に注目すべき画像特徴や局所領域に関する情報を与えることで，より多様な画像特徴の生成能力の獲得を促し，画像変換性能の向上を目指す．

In this study, we propose adversarial learning with a feedback mechanism from the discriminator as a new unsupervised learning framework to improve the quality of CBCT images. With conventional GANs, adversarial learning is performed based on the global discrimination results output by the discriminator, but with the proposed method, the discriminator is extended to output a probability map. A probability map is an image in which the probability that the input image belongs to the target image sets is calculated on a pixel basis. This map aims to obtain a local evaluation of the differences in image features found between the synthetic image and the target image. In addition, the proposed adversarial learning framework introduces a feedback mechanism that inputs the probability map output by the discriminator to the generator. By providing the generator with information about attention and image features, we aim to improve the image-translation performance by inducing the generator to acquire the ability to generate more diverse image features.

In our experiments, adversarial learning was conducted using planning CT images for 76 patients who underwent radiotherapy for prostate cancer and CBCT images taken on the day of radiotherapy. We investigated the image translation performance by comparing the proposed framework with the two existing methods. The contributions of this study are as follows.

\begin{itemize}
    \item A new adversarial learning scheme with a feedback mechanism
    \item Analysis of the impact of a probability map generated from the discriminator
    \item Application of the methods to improve the quality of CBCT images
\end{itemize}

\begin{figure}[t]
    \centering
    \includegraphics[width=11cm]{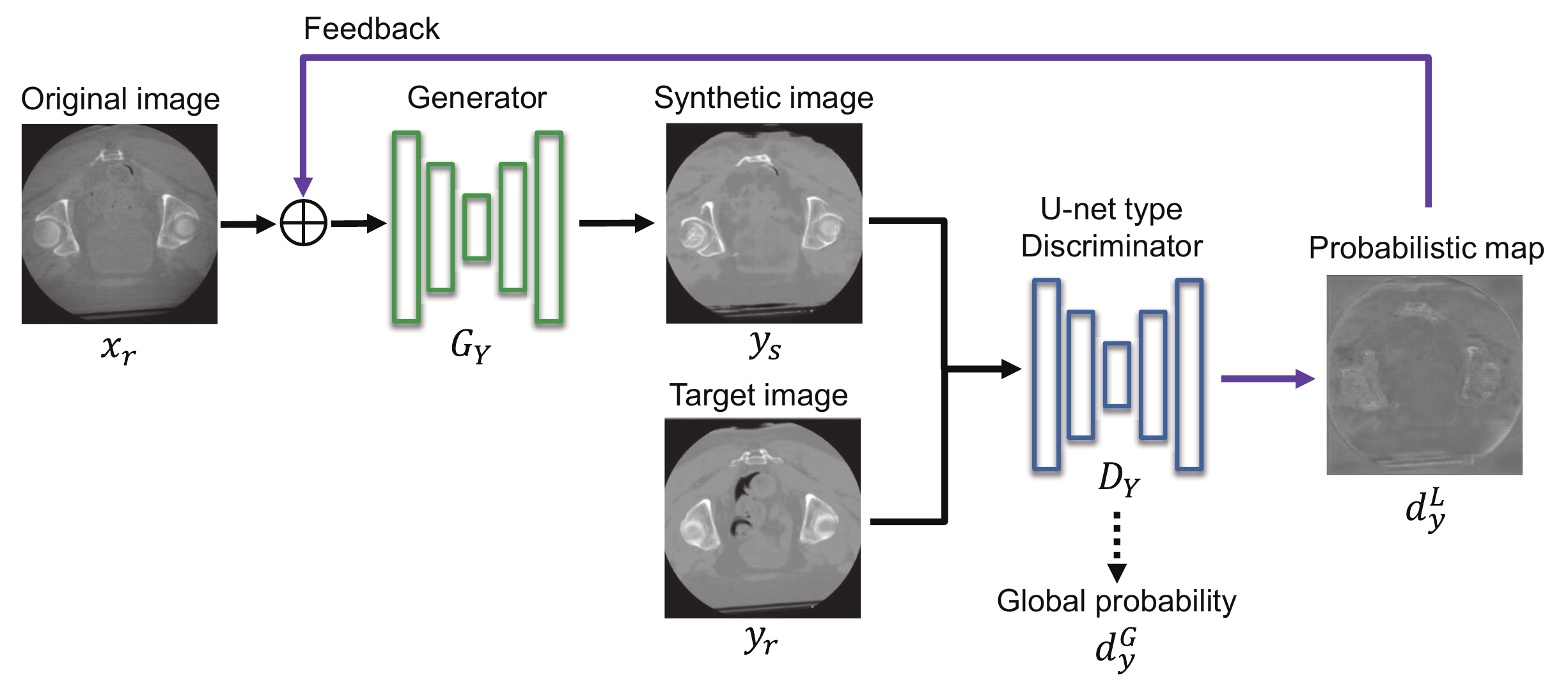}
    \caption{Adversarial learning with the proposed feedback mechanism. The original image and the probability map output from the U-net type discriminator are concatenated and input to the generator.}
    \label{fig:1}
\end{figure}

\section{Methods}

\subsection{Adversarial Learning with a Feedback Mechanism}
In a conventional GAN, the global discrimination result of the discriminator is calculated as a scalar between 0 and 1 for the synthetic image and is used only for evaluation in the loss function. However, this scheme does not provide sufficient translation performance for an image with different local characteristics. The proposed method extends both generator $G_Y$ and discriminator $D_Y$ in $X\rightarrow Y$ image translation ($X$: original CBCT image sets and $Y$: target CT image sets). It introduces a structure that feeds back the information on the attention and local image features in each image set accumulated by discriminator $D_Y$ during adversarial learning to generator $G_Y$. %This is aimed at improving translation performance for locally different image features.

Figure 1 shows the basic structure of the proposed network with the feedback mechanism. In this framework, a discriminator $D_Y$ is extended to output probability map $d^L_y$ in addition to the conventional global discrimination result $d^G_y$. A probability map is an image in which the probability that the input image belongs to the target image sets is calculated on a pixel basis. This map aims to obtain a local evaluation of the differences in image features found between the synthetic image and the target image. 

Next, the feedback mechanism provides the output of discriminator $D_Y$ as the input to generator $G_Y$. In our adversarial learning, rather than directly inputting original image $x_{r}$ into generator $G_Y$, probability map $d^L_y$ obtained from $D_Y$ in advance is concatenated with original image $x_{r}$ and input into generator $G_Y$. The details of the two extended networks, $D_Y$, and $G_Y$, are described in the next section. This feedback of information from the discriminator allows generator $G_Y$ to receive directly the local regions and image features on which the discriminator focuses. This is expected to imbue the network with the ability to generate various image features and improve image translation performance by adversarial learning. Finally, the flow of image translation using the trained networks can be organized as follows. 

\begin{description}
    \item [STEP1] \hspace{1mm} Original image $x_r$ is input into discriminator $D_Y$
    \item [STEP2] \hspace{1mm} Discriminator $D_Y$ outputs probability map $d^L_y$ in addition to global discrimination result $d^G_y$
    \item [STEP3] \hspace{1mm} Original image $x_r$ and probability map $d^L_y$ are combined and input into generator $G_Y$
    \item [STEP4] \hspace{1mm} Generator $G_Y$ outputs synthetic image $y_f$
\end{description}

\subsection{Structure of the Discriminator}
This section describes the structure of the extended discriminator. In this study, a U-net \cite{Unet} was employed for the discriminator $D_Y$ to output the probability. This U-net type discriminator newly outputs probability map $d^L_y$ for the input image in addition to the conventional global discrimination result $d^G_y$. Each pixel of the probability map contains the local discrimination result of the corresponding pixel in the input image.

When training the U-net type discriminator, synthetic image $y_f$ output from the generator is input along with the reference image $y_{r}$. When target image $y_{r}$ is input, the discriminator aims to output a probability map in which all pixels are 1.0, and when synthetic image $y_{f}$ is input, the discriminator aims to output a probability map in which all pixels are zero. This allows the discriminator to learn the image features of the target image set $Y$ and to output probability map $d^L_y$ that gives a low probability if there are image features in synthetic image $y_{f}$ that are locally different from those of image set $Y$. (Regarding the prediction phase, see Figure 4 in the supplemental document.) %Figure 2 shows the input and output during estimation. The probability map for original image $x_{r}$ is obtained by inputting original image $x_{r}$ instead of synthetic image $y_{f}$ or target image  $y_{r}$ during the estimation. %To match the output of U-net type discriminator $D_Y$ described above, 
The loss function used in training is extended to train the U-net type discriminator. Eq. (\ref{feedback loss 1}) and (\ref{feedback loss 2}) are defined to improve the accuracy of local as well as global discrimination.

\begin{align}
	\label{feedback loss 1}
	\mathcal{L}_{d^G_y} = \mathbb{E}_{y}||d^G_y(y_{r}) - 1||_1 + \mathbb{E}_{y}||d^G_y(y_{f}) - 0||_1
\end{align}
\begin{align}
	\label{feedback loss 2}
	\mathcal{L}_{d^L_y} = \mathbb{E}_{y}\left[\sum_{i,j}\left[||d^L_y(y_{r}) - \boldsymbol{1}||_1\right]_{i,j}\right] + \mathbb{E}_{y}\left[\sum_{i,j}\left[||d^L_y(y_{f}) - \boldsymbol{0}||_1\right]_{i,j}\right]
\end{align}
%ここで，$i$, $j$は画素の位置である．式(\ref{feedback loss 1})は従来と同様，U-net型識別器$D_Y$による大域的な識別結果$d^G_y$の精度を向上させるための損失関数であり，この損失を小さくするように学習が進むと，U-net型識別器$D_Y$は入力画像が目標画像であるか生成画像であるかを識別可能となる．式(\ref{feedback loss 2})は新たに追加した損失関数であり，同時に確率マップ$d^G_y$の精度を向上させることを要請する．本損失を小さくするように学習が進むと，U-net型識別器$D_Y$は入力画像の各画素が目標画像か否かを局所的に識別可能となる．以上の2つの損失の線形結合によって，U-net型識別器$D_Y$に対する損失関数を式(\ref{feedback loss 3})のように定義する．
where $i$ and $j$ are the positions of the pixels. Eq. (\ref{feedback loss 1}) is a loss function to improve the accuracy of global discrimination result $d^G_y$ given by U-net type discriminator $D_Y$, as in conventional cases. When the training proceeds to reduce this loss, U-net type discriminator $D_Y$ can discriminate whether the input image is the target or synthetic. Eq. (\ref{feedback loss 2}) is the newly added loss function. Concurrently, it seeks to improve the accuracy of probability map $d^G_y$. When learning proceeds to reduce this loss, $D_Y$ can locally discriminate whether each pixel in the input image is from the target image or not. The linear combination of the above two losses defines the loss function for U-net type discriminator $D_Y$ as in Eq. (\ref{feedback loss 3}).
\begin{align}
	\label{feedback loss 3}
	\mathcal{L}_{D_Y} = \mathcal{L}_{d^G_y} + \mathcal{L}_{d^L_y}
\end{align}
Here, the weight for linear combination is set to 1.0, with reference to the U-net GAN in \cite{Edgar20}. 

\begin{figure}[t]
    \centering
    \includegraphics[width=11.5cm]{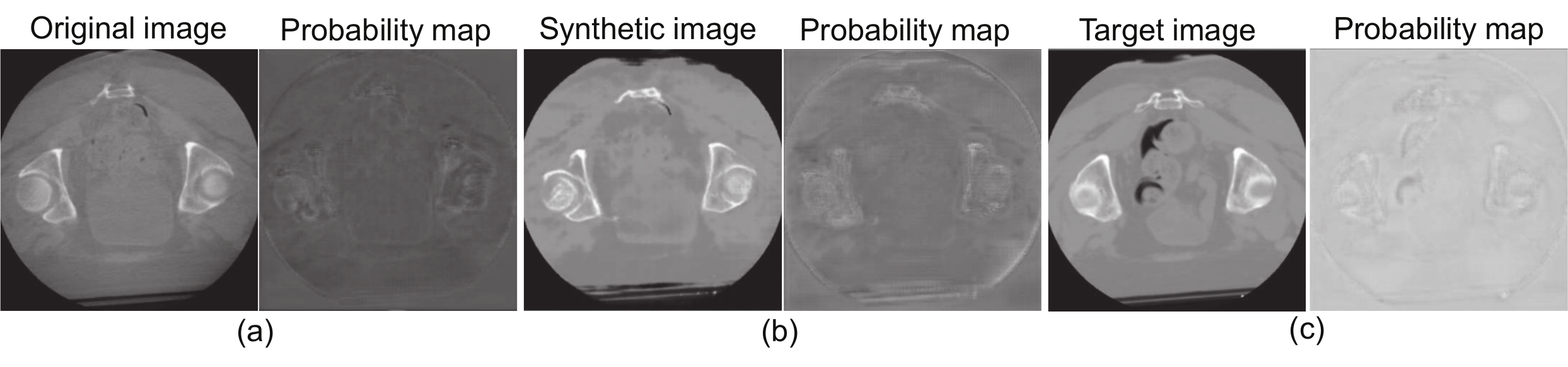}
    \caption{Probability maps output from the discriminator for different input images, (a) Original CBCT image and probability map, (b) Synthetic image and probability map, (c) Cropped target CT image and probability map.}
    \label{fig:2}
\end{figure}

Figure \ref{fig:2} shows an example of the probability maps obtained when different images are input to the trained discriminator. Each left image shows the input to discriminator $D_Y$, and the right image shows the respective probability map, which spatially represents the probability that the image features of each local region belong to image set $Y$. The probability map of (a) is mostly dark gray (nearly zero), indicating that the original image $x_{r}$, i.e., the input CBCT image, has many regions with features that differ from the CT image. Conversely, the probability map of (c) is mostly occupied by nearly white pixels, indicating that when the target image, CT image $y_{r}$, is input, the probability of them belonging to image set $Y$ is also high for the image features of the local regions calculated for each pixel. (b) shows the probability map when synthetic image $y_{f}$ is input during the training process. Compared to (a), there are light gray pixels, but compared to (c), there are still many pixels close to dark gray. This results from the trained discriminator $D_Y$ assigning a low probability to regions in the synthetic images that are different from the features in the image set $Y$. Thereby, the proposed image translation network aims to improve the translation performance of generator $G_Y$ by feeding back probability map $d^L_y$ output from the U-net type discriminator to generator $G_Y$.

\subsection{Structure of the Generator}
To utilize the feedback from the discriminator for image translation, original image $x_{r}$ and probability map $d^L_y$ are concatenated into the generator as a 2-channel image. In adversarial learning, probability map $d^L_y$ is utilized for the image translation of generator $G_Y$, and Eq. (\ref{feedback loss 7}) is defined as the loss function for generator $G_Y$ to promote the acquisition of the ability to generate more diverse image features.
\begin{align}
	\label{feedback loss 7}
	\mathcal{L}_{G_Y} = \mathbb{E}_{y}||d^G_y(y_{f}) - 1||_1 + \mathbb{E}_{y}\left[\sum_{i,j}\left[||d^L_y(y_{f}) - \boldsymbol{1}||_1\right]_{i,j}\right]
\end{align}
where the first term is the evaluation of the global discrimination result when synthetic image $y_{f}$ is input to discriminator $D_Y$, and the loss obtains an evaluation whereby, according to discriminator $D_Y$, the synthetic image belongs to target image set $Y$. The second term, which is the evaluation of the probability map when synthetic image $y_{f}$ is input to discriminator $D_Y$, is newly added. This term is designed as a loss to obtain an evaluation whereby the synthetic image belongs to image set $Y$ locally as well. By updating the weight parameters of the network with this loss function, generator $G_Y$ is encouraged to learn an image translation that effectively utilizes probability map $d^G_y$.

Combining the above discriminator $D_Y$ and generator $G_Y$ achieves adversarial learning using the probability map as feedback when translating image $X\rightarrow Y$ . The overall loss function is then defined by Eq. (\ref{feedback loss 11}), which is the linear sum of Eq. (\ref{feedback loss 3}) and (\ref{feedback loss 7}).
\begin{align}
	\label{feedback loss 11}
	\mathcal{L} = \mathcal{L}_{D_Y} + \mathcal{L}_{G_Y}
\end{align}

In the prediction or testing phase, a probability map  $d^L_y$ is obtained by inputting the original image $x_{r}$ to the discriminator $D_Y$ beforehand. As shown in Figure \ref{fig:2}, the probability map output from the trained discriminator $D_Y$ shows a low probability for regions that differ from the local features of image sets $Y$. This information map is utilized by generator $G_Y$ to capture and translate more diverse image features.

Although there are many differences in artifacts and inaccurate pixel values between CBCT image sets $X$ and CT image sets $Y$, they share common prior knowledge, such as the patient's anatomical structures. Therefore, we also trained the $Y\rightarrow X$ image translation network simultaneously for stable learning and convergence of the network parameters. The loss functions, $\mathcal{L}_{D_X}$ for the discriminator $D_X$, $\mathcal{L}_{G_X}$ for the generator $G_X$ and cycle consistency loss $\mathcal{L}_{cyc}$ were added to Eq. (5) in our feedback assisted adversarial learning. 

\section{Experiments}
We designed an experiment to verify how effective the proposed image translation framework is for improving the quality of CBCT images. The similarity between the synthetic images and the reference CT images was investigated to verify the translation performances of CycleGAN, U-net GAN, and the proposed method. We implemented our methods using Python 3.6 with a TensorFlow background. 

\subsection{Dataset}
In experiments, we used CT images ($512\times512$ pixel, 134-226 slices) taken for radiotherapy planning from 76 prostate cancer patients who underwent radiotherapy and CBCT images ($512\times512$ pixel, 48-93 slices) taken on the first day of radiotherapy. Differences arose between the two images due to different imaging conditions, such as the field of view (FOV), date and time of imaging, and the posture and condition of the patient. Therefore, the CT image was cropped by the cylinder shape (i.e., the same FOV setting as CBCT imaging), resampled, and aligned to the CBCT image through rigid registration. This pre-processing produced CT and CBCT image pairs ($512 \times 512$ pixels, 48-93 slices) for 76 subjects, in which anatomical structures and FOVs were registered except for the intra-patient differences, e.g., deformation in abdominal organs. 68 CBCT-CT image pairs were randomly selected as the training data, and the remaining eight were used as the test data. We used 4 for each training batch and 600 for the total number of training epochs. Adversarial training in this setting required 54 hours using a computer with a graphics processing unit (CPU: Intel Core i7-9900K, Memory: 64 GB, GPU: NVIDIA TITAN RTX). In the test, the CBCT image was used as the original image, the CT image was used as the reference, and the synthetic image was generated by $X \rightarrow Y$ translation. 

\subsection{Quantitative Evaluation Results}
The artifacts in CBCT images are mainly observed around the prostate. The region of interest (ROI) to be evaluated was set to $256 \times 256$ pixels and 20 slices, centered on the middle of the prostate. The range of CT values was determined to be [-300, 150], and the values were rounded to -300 for CT values below -300 and 150 for CT values above 150, for the evaluation to focus on the fine differences in CT values around the prostate. As indices for quantitative evaluation, the mean and standard deviation of the voxel values were employed because there were anatomical differences derived from different imaging conditions. Then, histograms of CT values within the evaluation range were created and compared. The correlation coefficient was calculated with the histogram of the reference image to evaluate how close the distribution was to the CT values of the reference image.

Table \ref{table:1} lists quantitative evaluation results. The values of the proposed method are closer to the reference image than those of the conventional methods. From the above quantitative evaluation using statistical indices, the proposed method is considered to be more effective in improving the quality of CBCT images than conventional methods.
\begin{table}[t]
	\centering
	\caption{Quantitative evaluation results} 
	\label{table:1}
\scalebox{0.9}{    	
	\begin{tabular}{c c c c c c} \hline
		{}&{Original}&{CycleGAN}&{U-netGAN}&{Proposed}&{Reference}\\\hline 
		Mean $\pm$ SD [HU] & $-67.2 \pm 9.9$ & $-1.19 \pm 82.3$ & $14.7 \pm 77.8 $ & $12.6 \pm 76.6$ & $9.07 \pm 80.1 $ \\
		Correlation coefficient & 0.185 & 0.859 & 0.890 & 0.930 & 1.00\\ \hline
	\end{tabular}
}
\end{table}

\subsection{Comparison of Synthetic Images}
\begin{figure}[tb]
    \centering
    \includegraphics[width=12cm]{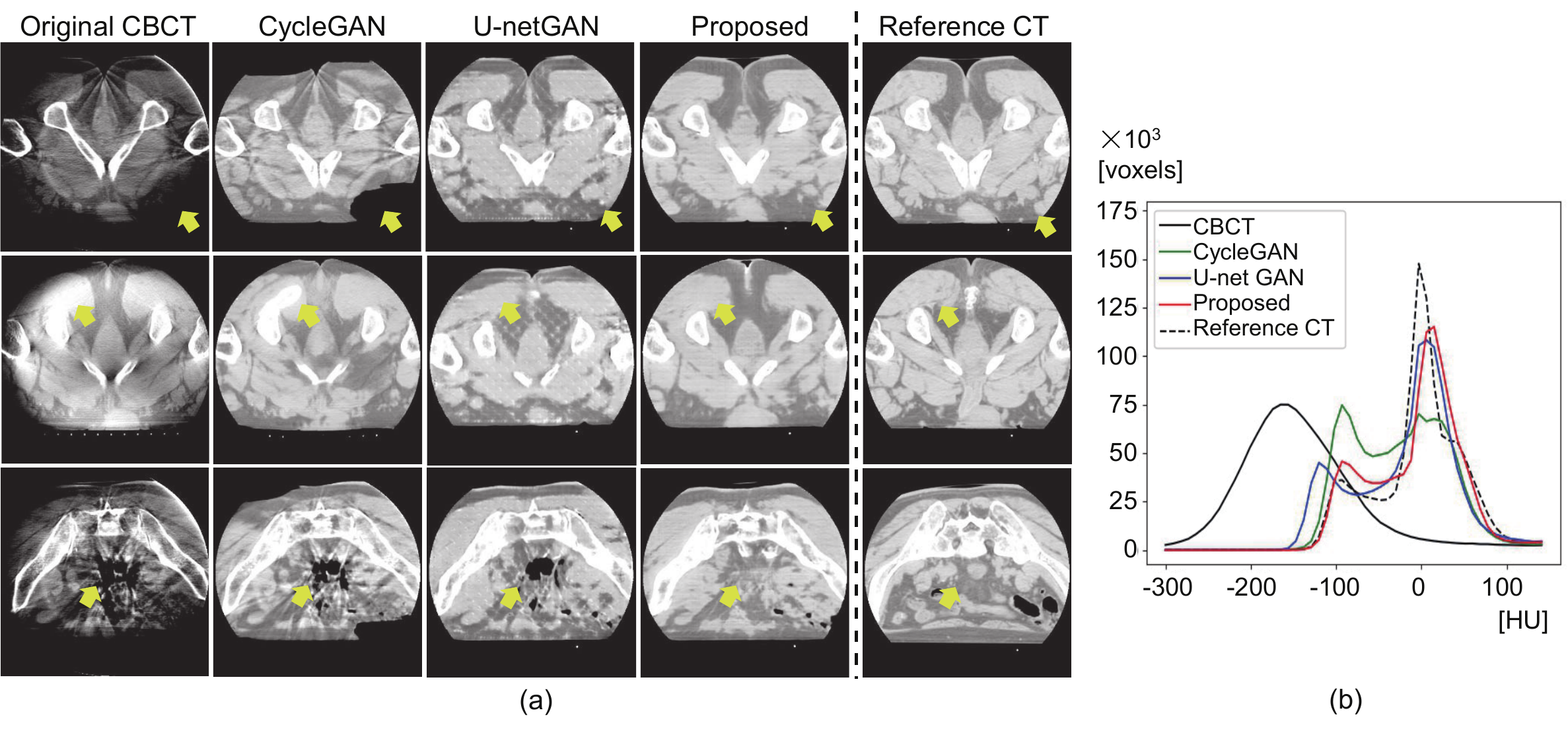}
    \caption{Comparison of synthetic images from the proposed method and conventional methods, (a) Visual comparison, (b) Histograms of synthetic image CT values.}
    \label{fig:3}
\end{figure}

Figure \ref{fig:3}(a) shows three examples of the synthetic images generated by the different image translation methods. (Please see the supplemental document for higher resolution examples and the learning process.) In the first case, as indicated by the arrows in the image, CycleGAN's image translation deformed the body outline but the proposed method enabled image translation while maintaining the body outline. In addition, the original CBCT image shows an overall lower CT value than the CT image, and the entire image is dark, while the proposed method improves the visual brightness to almost the same level as the reference CT image. The arrows in the second and third cases indicate a region where the CBCT image contains a strong artifact. CycleGAN and U-netGAN failed to translate or wrongly corrected the region, and residual artifacts remain. In contrast, the proposed method successfully translated the region. This case suggests that the proposed method effectively recovers the underlying shape of organs even during the translation of areas with strong artifacts. 

Figure \ref{fig:3}(b) shows a histogram of the one test case. The solid black line is the histogram of the CBCT image before translation, and the dashed black line is that of the reference CT image. The red, blue, and green solid lines, respectively, are the results of translations using the proposed method and the conventional methods, such that the CT value distribution approaches the black dashed line. As the figure shows, the red line, which is the proposed method, is closer to the shape of the black dashed line. The proposed method is better able to transform the image than the conventional methods; therefore, the distribution of CT values is closer to the reference image.

\section{Conclusion}
In this study, we proposed adversarial learning with a feedback mechanism from the discriminator as a new unsupervised learning framework to improve the quality of CBCT images. The experiments using 76 paired CT-CBCT images showed that the proposed framework could capture more diverse image features than conventional methods and produced synthetic images with pixel values close to the reference image and a correlation coefficient of 0.93. 

%was conducted to verify the effectiveness of the proposed method for improving the quality of CBCT images. Quantitative evaluation of the trained model using test data showed that the average CT value of the synthetic images was 12.6 ± 76.6, and the correlation coefficient with the CT value distribution of the reference image was 0.930, indicating that image translation based on the image features of the reference image was achieved better than with conventional methods.

\end{document}